# Anomalous magnetoresistance in the spinel superconductor $LiTi_2O_4$

K. Jin[1,2,3], G. He[1], X. Zhang[3,4], S. Maruyama[4], S. Yasui[4], R. Suchoski[4], J. Shin[4], Y. Jiang[3], H.S. Yu[1], L. Shan[1], R. L. Greene[3†], I. Takeuchi[4†]

[1]Beijing National Laboratory for Condensed Matter Physics, Institute of Physics, Chinese Academy of Sciences, Beijing 100190, China
[2]Collaborative Innovation Center of Quantum Matter, Beijing, 100871, China
[3]Center for Nanophysics and Advanced Materials and Department of Physics, University of Maryland, College Park, Maryland 20742, USA
[4]Department of Materials Science and Engineering, University of Maryland, College Park, Maryland 20742, USA

email: kuijin@iphy.ac.cn

**Transition-metal oxides offer an opportunity to explore unconventional superconductors, where the superconductivity (SC) is often interrelated with novel phenomena such as spin/charge order, fluctuations, and Fermi surface instability (1-3). $LiTi_2O_4$ (LTO) is a unique compound in that it is the only known spinel oxide superconductor. In addition to electron-phonon coupling, electron-electron and spin fluctuation contributions have been suggested as playing important roles in the microscopic mechanism for its superconductivity (4-8). However, the lack of high quality single crystals has thus far prevented systematic investigation of their transport properties (9). Here, we report a careful study of transport and tunneling spectroscopy in epitaxial LTO thin films. In the superconducting state, the energy gap was found to decrease as a quadratic function of magnetic field. In the normal state, an unusual magnetoresistance (MR) was observed where it changes from anisotropic positive to isotropic negative as the temperature is increased. A constant charge carrier concentration without any abrupt change in lattice parameters as a function of temperature suggests that the isotropic MR stems from the suppression of spin scattering/fluctuations, while the anisotropic term originates from an orbital contribution. These observations point to an important role strong correlations play in this unique superconductor.**

$LiTi_2O_4$ (LTO) is the only known superconducting transition-metal oxide with a spinel crystal structure, and its transition temperature ($T_c$) is 11 K (9). It was discovered in early 1970's by Johnston et al (10), and its $T_c$ can be described by band-structure calculations using the McMillan's formula with a weak electron-phonon coupling constant ($\lambda_{el-ph}$~0.6) (4, 11). However, the specific-heat measurements have pointed to the presence of an enhanced density of states or an equivalently larger coupling constant with $\lambda$ ~1.8 (7, 8). Other measurements including nuclear magnetic resonance (6), point contact spectroscopy (12), and resonant inelastic soft-x-ray scattering (13) have revealed the significance of *d-d* electron correlations and short-range spin ordering (14). Such results are suggestive of existence of nontrivial electron-electron correlations and spin fluctuations in this system. The development of understanding of this system has being hampered by the lack of sample reproducibility and the availability of single crystals or high quality thin films (9, 15, 16). Recently, high quality epitaxial LTO thin films successfully grown by pulsed laser deposition (PLD) have been demonstrated (17, 18), thus opening the door to possibilities for systematic experiments on LTO. Here, we present results of transport and tunneling studies on single crystalline-like epitaxial LTO thin films. The suppression of the superconducting energy gap as a quadratic function of magnetic field, and an anomalous crossover of magnetoresistance from anisotropic positive to isotropic negative with increasing temperature have been observed for the first time.

(00*l*)-Oriented LTO thin films were epitaxially grown on (00*l*)-oriented $MgAl_2O_4$ (MAO) substrates by PLD. Our LTO films consistently display the $T_c$ of 11 ± 0.25 K with narrow transition widths of less than 0.5 K. We have found that different films display different residual resistivity ratios (*RRR*). The films were patterned into Hall-bars to carry out Hall and normal resistivity measurements. Tunneling spectroscopy was performed where Pt-Ir tips were used to make point contacts in the *c*-axis direction (perpendicular to the film plane) of LTO crystals (19).

Fig. 1(a) shows the resistivity versus temperature curves for samples with two different *RRR* ratios: it is 6.25 and 3 for samples L1 and L2, respectively. L1 and L2 have similar resistivity values at the room temperature and the same $T_c$. The normal state resistivity of both samples can be fitted to a curve consistent with the Fermi liquid behavior, $\rho = \rho_0 + AT^2$ (gray lines) from 40 K to 120 K with residual resistivity ($\rho_0$) of ~71 µΩ cm for L1 and ~166 µΩ cm for L2. By sweeping the magnetic field perpendicular to the film surface ($B \perp ab$ plane) at fixed temperatures, the Hall resistivity, i.e. $\rho_{xy} = \frac{E_y}{j_x} = \frac{V_y \cdot t}{I_x}$, was extracted, and they are plotted in Fig. 1(b), where $V_y$ is the Hall voltage found by subtracting the transverse voltage in negative field from that in positive field, and *t* is the thickness of the film. In the normal state, the Hall resistivity is always proportional to the magnetic field, positive and temperature independent, strongly suggesting the presence of one type of charge carrier (holes) and a simple electronic band structure. The charge carrier concentration is calculated assuming a parabolic band structure, i.e., the Hall coefficient $R_H = \frac{1}{ne}$, and L1 and L2 have almost the same hole concentration of ~ $3 \times 10^{22}$ cm$^{-3}$ [Fig. 1(c)], indicating that the different *RRRs* are caused by the difference in mobility values.

The point contact measurements were carried out on L1 before it was patterned into a Hall bar for transport measurements. As described by Blonder, Tinkham, and Klapwijk (BTK) (20), the tunneling regime is achieved for Z > 1, where Z represents the tunneling barrier height and the Fermi velocity mismatch (21). The differential conductance spectrum shows a clear temperature and field dependent coherence peak. The normalized differential conductance spectra with and without applied magnetic field are shown in Fig. 2(a), 2(c) and Fig. S2 as a function of bias voltage (Figure S1 shows a raw conductance curve taken at 2 K). The normalized experimental curves (black points) were fitted (red lines) using a modified BTK model with a complex energy $E' = E + i\Gamma$ (22). The broadening $\Gamma$ term, which takes into account sample inhomogeneity or a finite quasi-particle lifetime by scattering, is temperature independent in zero field, but application of magnetic field was found to lead to an additional pair breaking factor (23), which in effect is akin to increasing $\Gamma$ (24).

Several key points could be made from the fitting to our tunneling spectra. *First*, the Z value of our Pt-Ir/LTO junction is ~ 2.4, which is independent of temperature and field. *Secondly*, zero-field spectra give a constant $\Gamma$ of ~ 0.94 meV (Fig. S3), but data in field have to be fitted with an increasing $\Gamma$ as the field is increased (Fig. S4). *Thirdly*, the temperature dependence of the superconducting energy gap can be fitted well with the BCS theory [Fig. 2(b)], and the observed $2\Delta_0/k_BT_c = 4$ ($\Delta_0 = 1.93$ meV) is consistent with previous reports (8, 12), indicating that LTO is a medium-coupling BCS superconductor. Moreover, a simple relation of $\Delta(B, T)/\Delta(0T, T) \sim -[B/B_{c2}(T)]^2$, can be used to scale the field dependent energy gap at different temperatures, e.g. $T = 2, 6, 10$ K [Fig. 2(d)]. As a result, the $B_{c2}$ can be extracted from the point contact spectra, and it is $B_{c2}(2K) \sim 16$ T.

We also employed a two-channel method derived from the BTK model to fit our experimental data (25). In this method, the pair-breaking effect by field is considered as a normal channel (N) superposed onto the superconducting channel (S). Assuming the differential conductance $G(h) = h^\alpha G_N + (1 - h^\alpha)G_S$ with $h = B/B_{c2}$, the normalized differential conductance in field should obey the polynomial form $\frac{G(h)}{G_N} = h^\alpha + (1 - h^\alpha)\frac{G_S}{G_N}$, where $\frac{G_S}{G_N}$ is obtained from the BTK model with $\Gamma$ fixed instead of increased. The best fitting requires $\alpha = 2$, consistent well with that in other superconducting systems like Nb, Mo$_3$Sb$_7$, and Dy$_{0.8}$Y$_{0.2}$Rh$_4$B$_4$ (25). In this case, the relation, $\Delta(B) \sim -B^2$, is also held as seen in Figure 2(d). However, we realize that so far no theory can account for such simple relation. When B is close to $B_{c2}$, the filed dependent gap could be analytically expressed as $\overline{|\Delta|}^2 \sim \frac{B_{c2} - B}{\Psi_2(\frac{1}{2} + \frac{A}{2\pi k_B T})}$, where $\overline{\Delta}$ represents the averaged energy gap taking into account the effect of vortices, and $\Psi_2$ is the first derivative of the digamma function with $A$ the pair breaking parameter proportional to $B$ when $B \perp ab$ plane (26, 27). But, even in this simplified case, pure quadratic field dependence of superconducting energy gap could not be deduced. Future theoretical effort is needed to understand the above relation and also the universal $\alpha$ value.

From the above transport and tunneling data, we were able to extract key parameters for LTO. The Ginzburg-Landau coherence length, the size of the vortex core in type II superconductor is estimated to be $\xi_{GL} = (\frac{\Phi_0}{2\pi B_{c2}})^{1/2} = 4.47$ *nm*. The mean free path of $l = 1.8$ *nm* is deduced from the Drude model $\rho_0 = \frac{\hbar k_F}{ne^2 l}$, where the Fermi wave-vector $k_F = (3\pi^2 n)^{1/3} = 0.96$ Å$^{-1}$. Since $l <$

$\xi_{GL}$, the BCS coherence length of $\xi_{BCS} = 14.9\ nm$ is calculated from the dirty limit relation, $\xi_{GL} = \frac{0.855 \times (\xi_{BCS} l)^{1/2}}{(1-\frac{T}{T_c})^{1/2}}$. We then calculate the Fermi velocity from the formula, $\xi_{BCS} = \hbar v_F / \pi \Delta$, and arrive at an effective mass of $m^*/m_0 = 8.11$ (with $m_0$ being the free electron mass), and the density of states at the Fermi level is found to be $N(E_F) = 0.96$ states/eV atom.

We compared these parameters to those obtained from previous reports on LTO. Only $T_c$ and $\xi_{GL}$ values have been previously reported on thin films (17), and the other quantities were from the magnetic susceptibility (15), Andreev reflection (12), and specific heat (8) measurements on polycrystalline samples. As seen in Table 1, the values obtained in the present work are consistent with those from previous reports. Note that in polycrystalline samples, the grain boundaries prevent accurate calculations of the electric transport measurements due to boundary scattering, but the parameters could be extracted from heat capacity and susceptibility data. These values indicate that the nearly free electron model can capture the main physics of the LTO system.

In the normal state, however, we observe an anomalous MR behavior. Figure 3(a) illustrates the field dependent MR in sample L1 when $B \perp ab$ plane. The MR gradually decreases from positive to negative with increasing temperature, and a crossover is observed at ~ 40 K. In the inset of Figure 3(a), the MR is plotted as $B^2$. The positive MR is proportional to $B^2$ whereas the negative MR is not. In the case of $B \perp ab$ plane, both orbital and spin effects can contribute to MR, and thus, in order to discern the different contributions, additional measurements were carried out where the field was also applied in the film plane with *B // I* (parallel to current) and *B* $\perp$ *I* (normal to current). In Fig. 3(b), the temperature dependence of MR is plotted for both L1 (gray symbols) and L2 (red symbols). The negative MR is independent of the field directions, but the positive MR displays strong anisotropy, i.e. MR ($B \perp ab$ plane) > MR (*B //* ab-plane, *B* $\perp$ *I*) > MR (B // ab-plane *// I*). Moreover, MR changes its sign when *B // I* as seen in Figure 4(a). We also notice that ρ(T) in zero field deviates from the Fermi liquid behavior below 40 K, as seen in Fig. 4(b). The discrepancy between the experimental data and the fitting curve, i.e. $\delta\rho = \rho_{xx}(0T) - \rho_0 + AT^2$, diverges as – $lnT$ and saturates at lower temperatures. The inverse magnetic susceptibility data were also measured [Fig. 4(c)], and a notable contribution to the susceptibility (corresponding to a drop in $\chi^{-1}$) was also observed below ~ 40 K. The above data collectively point to the presence of a magnetic transition with a characteristic temperature $T_{ch}$ of ~ 40 K.

MR can in general result from charge, orbital, or spin interactions, as well as interactions among the three (2). First of all, the constant Hall coefficient and smooth evolution of the *c*-axis lattice parameter (*SI* text) excludes the presence of charge density waves, which normally affects the Hall coefficient (28). Secondly, the anisotropic and $B^2$-dependent positive MR as though it is due to a conventional orbital contribution. Thirdly, the negative MR is isotropic which can be a spin scattering effect.

It is intriguing that MR changes its sign at $T_{ch}$ when $B \parallel I$. In antiferromagnetic metals, a sign change in MR is expected theoretically at the Neel temperature due to the *s-d* electron interaction (29). However, this idea is based on the long range antiferromagnetism, and the change in sign from negative to positive is expected to be discontinuous contrary to the observed behavior in LTO here. The LTO system has a frustrated Ti sublattice containing equal numbers of $Ti^{3+}$ and $Ti^{4+}$, and thus a long range antiferromagnetic ordering is unfavorable (14). Antiferromagnetic spin fluctuations have not been directly observed in LTO. The previous work on polycrystals (15) reported that the susceptibility was a Curie-Weiss-like at low temperatures, and attributed it to the presence of local paramagnetic $Ti^{3+}$ moments. For our thin film samples, a similar drop in $\chi^{-1}$ at low temperatures was also found [Fig. 4(c)], but we cannot tell whether it is Curie-Weiss-like or not due to the fact that it is difficult to cleanly separate a possible signal from the substrate. The emergence of short-range antiferromagnetic ordering has been reported in V-doped LTO by nuclear magnetic resonance (30) and in $LiV_2O_4$ by inelastic neutron scattering (31). Perhaps, a new theoretical study is in order as to whether local spin scattering or fluctuations could result in a smooth change of the sign of the MR.

Another possible explanation of the present MR behavior is grain boundary scattering. If the mean free path (*l*) is comparable to the grain size (*z*), the MR may change its sign. In this case, the electric transport will include intrinsic intra-grain transport as well as scattering from grain boundaries. The mean free path of the "clean" sample (L1) is 1.84 nm in the zero temperature limit, which is only about 2 unit cells of the LTO lattice. Therefore, it is unlikely that the grain boundary effect is playing a dominant a role. Instead, it is perhaps a breakdown of the Ioffe-Regel rule which takes place at high temperatures, as in the case of cuprates (32). Other effects such as suppression of superconducting fluctuations and spin-orbit coupling may also contribute to a positive MR (33). The former needs a wide temperature range of superconducting fluctuations which seems unlikely. The latter is beyond the immediate scope of the present discussion, and it would require further experiments in future.

In conclusion, the fundamental properties of our LTO films are consistent with previous work on polycrystalline samples: we observe one type of charge carriers and a $\frac{2\Delta}{k_B T_c}$ value consistent with a medium-coupling BCS superconductor. We find that the main physics could be captured by the nearly free electron model. However, we find two new and distinct features in this system. In the superconducting state, a scaling law, $\bar{\Delta} \sim B^2$ at $T < T_c$, is extracted from our point contact spectra. This simple relation has not been predicted by any theory as yet. In the normal state, anomalous MR which crosses over from isotropic negative MR to anisotropic positive MR with decreasing temperature is observed. The isotropic MR stems from spin effects whereas the anisotropy originates from orbital contributions. These results confirm that electron-electron correlations/spin fluctuations are crucial to the understanding of the behavior of this unique system.

**Acknowledgements**: We thank Y. Dagan, Y. Nakajima, J. Paglione, P. Dai, G. Zheng, and K. Liu for useful discussions. This research was supported by the National Science Foundation (1104256) and also AFOSR-MURI (FA9550-09-1-0603). K.J. was also supported by the Strategic Priority Research Program of the Chinese Academy of Sciences (XDB07000000).

**Table 1. Physical parameters of LiTi$_2$O$_4$ obtained in present work and comparison to values from previous work.**

| | $T_c$(K) | $\Delta$(meV) | $\xi_{BCS}$(nm) | $\xi_{GL}$(nm) | n (cm$^{-3}$) | $m^*/m_0$ | $l$(nm) | $v_F$(m/s) | $N(E_F)$ |
|---|---|---|---|---|---|---|---|---|---|
| Present | 11±0.25 | 1.93±0.01 | 14.9 | 4.47 | $3\times10^{22}$ | 8.11 | 1.84 | $1.37\times10^5$ | 0.96 |
| Previous | 11.5±0.5 | 1.9 | - | 4.1-4.6 | $1.35\times10^{22}$ | 9.4 | 3.2 | - | 0.97 |

The quantities in the present work were obtained from the transport and tunneling results on sample L1. Since most of the earlier studies were carried out on polycrystalline samples, the parameters such as carrier density (n), effective mass (m$^*$), and density of states (N(E$_F$): states/eV.atom) were from previous magnetic susceptibility measurements using nearly free electron approximations (15). The mean free path ($l$) was calculated from the specific heat data (8). We compare the Ginzburg-Landau coherent length ($\xi_{GL}$) to the value reported by another thin film study (17), and the superconducting energy gap ($\Delta$) to the value obtained from an Andreev study on polycrystalline samples (12).

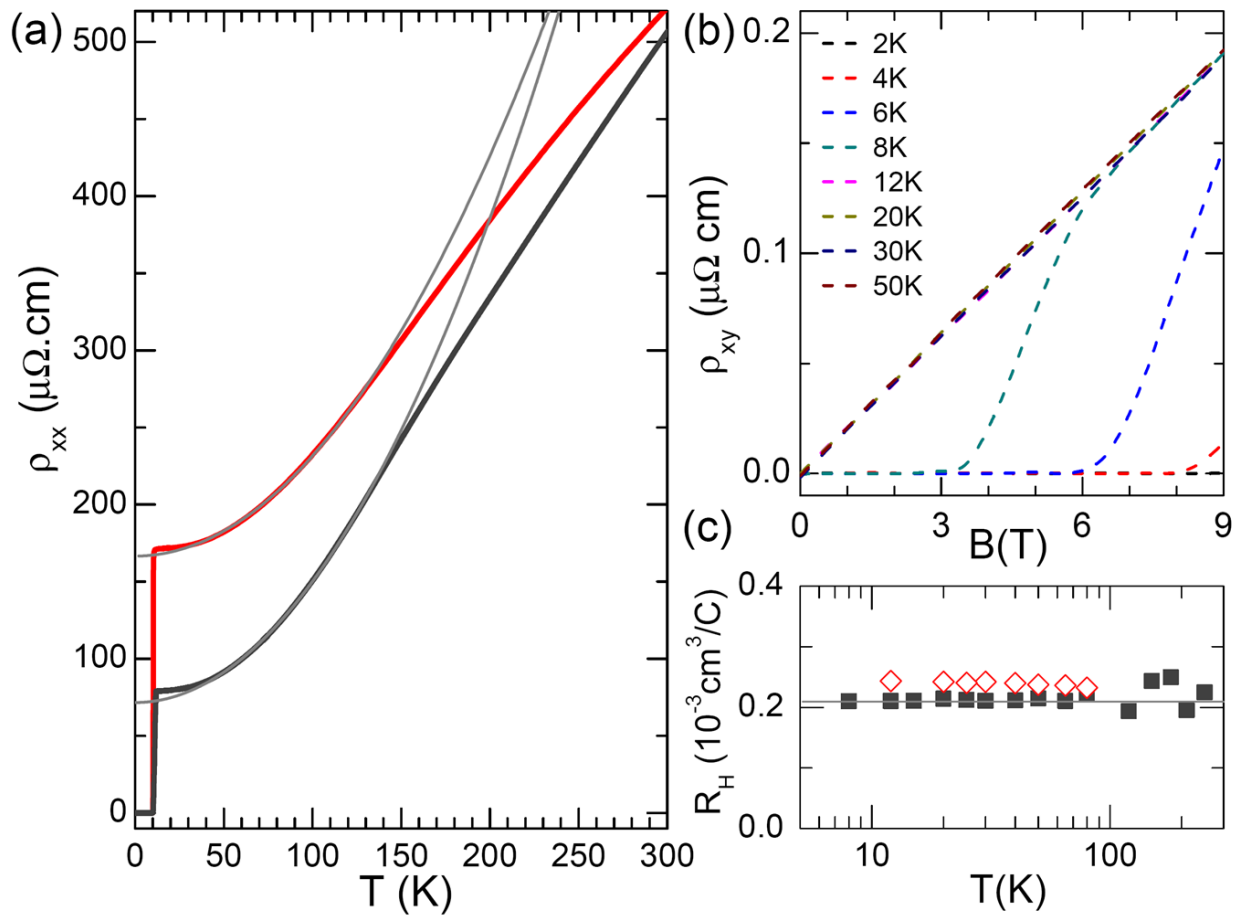

**Figure 1 Resistivity, Hall resistivity and Hall coefficient of $LiTi_2O_4$ thin films. (a)** $\rho_{xx}(T)$ of two samples are shown here. They have similar resistivity (~500 μΩ cm) at room temperature, and the same $T_c$ of 11 K, but different residual resistivity ratios (*RRR*): *RRR*~6.25 for sample L1 (black curve) and ~ 3 for the sample L2 (red curve). The resistivity curve from 40 to 120 K can be fit by $\rho = \rho_0 + AT^2$ (gray lines). **(b)** The Hall resistivity is proportional to the magnetic field at different temperatures and all the $\rho_{xy}(B)$ curves overlap in the normal state (only data on sample L1 are shown here), suggesting a simple one band structure and a temperature independent Hall coefficient. Note that at very low temperatures, the magnetic field is not sufficient to suppress the superconductivity ($B \perp$ ab plane). **(c)** Hall coefficient versus temperature for both samples (L1: solid symbols. L2: open symbols). Though the two samples show different *RRR*, their Hall coefficient values are very close to each other. Assuming $R_H =1/(ne)$, we find a hole concentration of ~$3\times10^{22}$ cm$^{-3}$ which is almost constant over the entire measured temperature range.

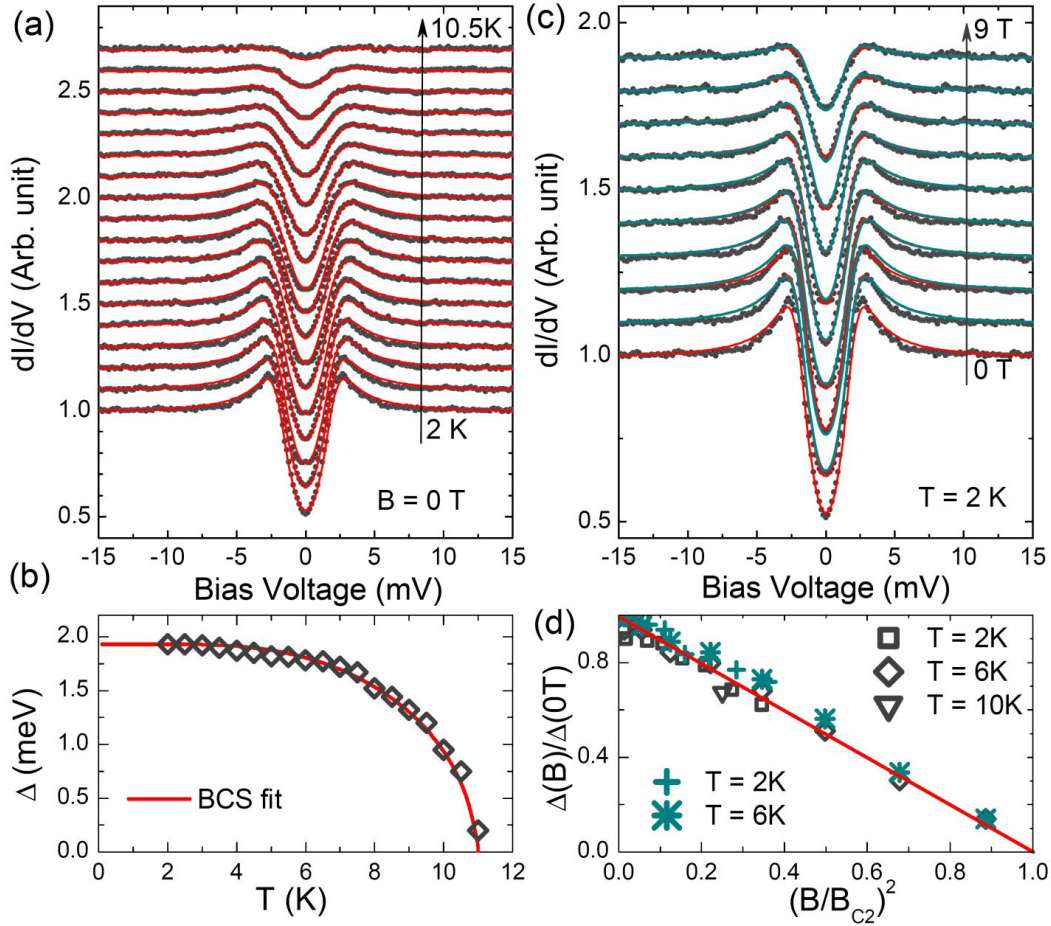

**Figure 2 Temperature and field dependent tunneling spectroscopy curves and the superconducting energy gap of LTO films.** **(a)** Normalized differential conductance versus bias voltage from 2 to 10.5 K ($\Delta T$ = 0.5 K) in zero field. Experimental data are fitted with a modified BTK model with a constant broadening Γ term (red lines), and are vertically shifted above the 2 K curve for clarity. **(b)** Temperature dependent energy gap values, $\Delta(T)$, are obtained from the BTK fits: $2\Delta/k_BT_c$ = 4 is obtained, indicating a medium-coupling BCS-like superconductor. **(c)** Normalized differential conductance versus field from 0 to 9 T ($\Delta B$ = 1 T, $B \perp ab$ plane). Experimental data (gray circles) are fitted with a modified BTK model with an increasing Γ as $B$ is increased (red lines, see *SI*), and also with the two-channel model (cyan lines). Data in fields are vertically shifted. **(d)** Normalized energy gap [$\Delta(B)/\Delta(0T)$] decreases as $(B/B_{c2})^2$ in the superconducting state, and can be scaled for different temperatures. In-field measurements were carried out at $T$ = 2, 6 and 10 K (open symbols extracted from modified BTK fittings with an increasing Γ and cross symbols with the two-channel model).

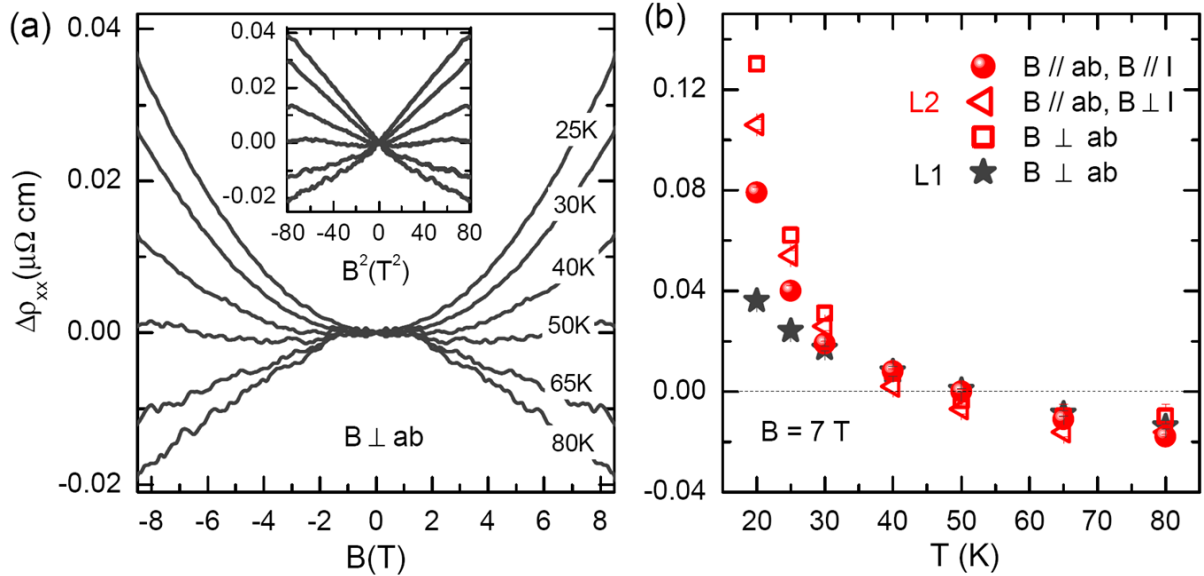

**Figure 3 Temperature dependence of magnetoresistivity, $\Delta\rho_{xx} = \rho_{xx}(B) - \rho_{xx}(0T)$.** **(a)** The transverse magnetoresistivity with $B \perp$ ab plane, changes from negative to positive as temperature is decreased. The crossover temperature is ~ 40 K. The positive magnetoresistivity is proportional to $B^2$ as seen in the inset, whereas the negative MR is not. **(b)** Magnetoresistivity at 7 T plotted against the temperature. The dark and red symbols are for samples L1 and L2, respectively. The magnetic field was applied along three different directions, i.e., $B$ // $ab$-plane ($B$ // I, $B \perp$ I) and $B \perp ab$ plane. The negative magnetoresistivity is isotropic, whereas the positive magnetoresistivity is anisotropic.

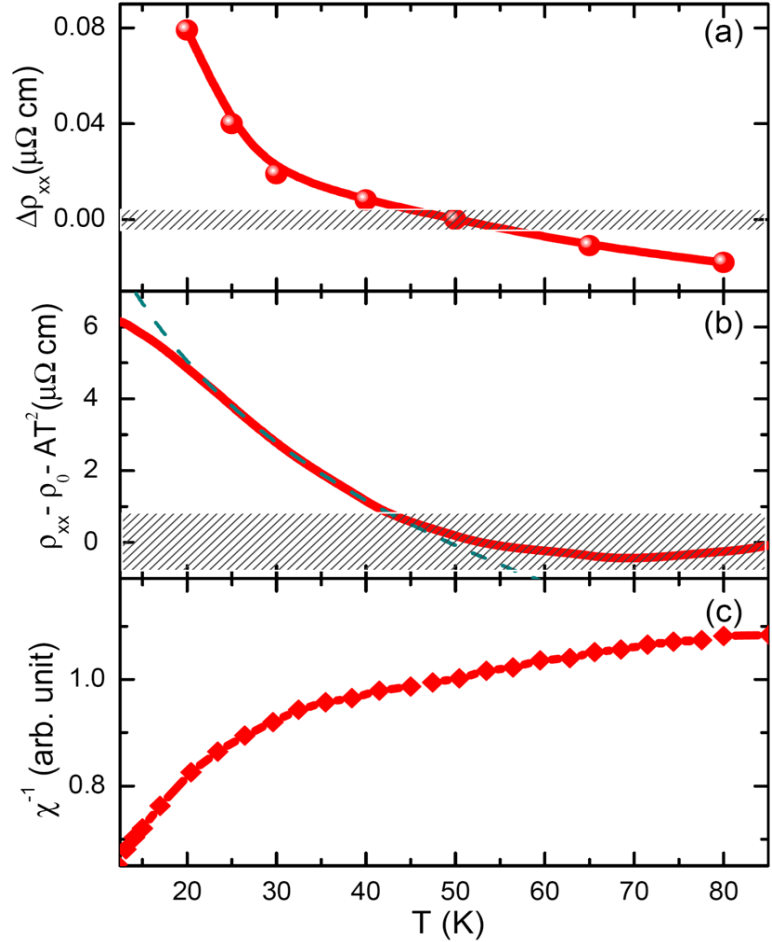

**Figure 4 Magnetoresistivity, resistivity deviation, and inverse susceptibility of the LiTi$_2$O$_4$ film (L1). (a)** Temperature dependent longitudinal magnetoresistivity ($B$ // $ab$-plane, $B$ // $I$), $\Delta\rho_{xx} = \rho_{xx}(B) - \rho_{xx}(0T)$, is replotted to determine the crossover temperature (~ 40 K) [also see Figure 3 (b)]. **(b)** ρ$_{xx}$(T) in zero field can be fitted by $\rho = \rho_0 + AT^2$ (Fermi liquid) as seen in Figure 1(a). But, it deviates from the Fermi liquid behavior at lower temperatures. If we define the deviation as $\delta\rho = \rho_{xx} - (\rho_0 + AT^2)$, then it satisfies $\delta\rho \sim -\ln T$ (dashed line) below 40 K. **(c)** χ$^{-1}$(T) shows a drop below 40 K, indicating an additional contribution to the susceptibility at low temperature, attributed to the presence of local paramagnetic Ti$^{3+}$ moments in a previous work (15). The shadow areas in (a) and (b) represent error bars in our measurements.